\documentclass[iop,numberedappendix,twocolappendix,twocolumn]{emulateapj} 

\usepackage{latexsym}   
\usepackage{amsfonts}   
\usepackage{amstext}    
\usepackage{amsmath,amssymb}
\usepackage{mathtools}
\usepackage{bm}
\usepackage{MnSymbol} 
\usepackage{mathrsfs}
\usepackage{hyperref}
\hypersetup{
  colorlinks   = true, 
  urlcolor     = blue, 
  linkcolor    = blue, 
  citecolor   = blue 
}

\begin{document}
 
\title{\bf Formal Solutions for Polarized Radiative Transfer\\ II. High-order Methods}
\author{Gioele Janett\altaffilmark{1,2}, Oskar Steiner\altaffilmark{1,3}, Luca Belluzzi\altaffilmark{1,3}}
\email{gioele.janett@irsol.ch}

\affil{$^1$ Istituto Ricerche Solari Locarno (IRSOL), 6605 Locarno-Monti, Switzerland\\
$^2$ Seminar for Applied Mathematics (SAM), ETH Zurich, 8093 Zurich, Switzerland\\
$^3$ Kiepenheuer-Institut f\"ur Sonnenphysik (KIS), D-79104 Freiburg i.~Br., Germany}

\begin{abstract}
When integrating the radiative transfer equation for polarized light, the necessity of high-order numerical methods is well known. In fact, well-performing high-order formal solvers enable higher accuracy and the use of coarser spatial grids. Aiming to provide a clear comparison between formal solvers, this work presents different high-order numerical schemes and applies the systematic analysis proposed by \citet{janett2017}, emphasizing their advantages and drawbacks in terms of order of accuracy, stability, and computational cost. 
\end{abstract}
\section{Introduction}\label{sec:sec1}
The transfer of partially polarized light is described by the radiative transfer equation
\begin{equation}
  \frac{\rm d}{{\rm d} s}\mathbf I(s) 
  = -\mathbf K(s)\mathbf I(s) + \boldsymbol{\epsilon}(s)\coloneqq \mathbf F(s,\mathbf I(s))\,,
\label{eq:RTE}
\end{equation}
where $s$ is the spatial coordinate measured along the ray under consideration, $\mathbf{I}$ is the Stokes vector, $\mathbf{K}$ is the propagation matrix, and $\boldsymbol{\epsilon}$ is the emission vector. For notational simplicity, the frequency dependence of the quantities is not explicitly indicated. Equation~\eqref{eq:RTE} is a system of first-order coupled inhomogeneous ordinary differential equations for which analytical solutions are available for a few simple atmospheric models only \citep{semel1999,lopez1999b}, which explains the necessity for a numerical approach. Therefore, the ray path is discretized through a spatial grid $\{s_k\}\;(k=0,\dots,N)$, where the index $k$ increases along the propagation direction. Assuming $\mathbf F$ to be Riemann-integrable in the interval $[s_k,s_{k+1}]$, one integrates Equation~\eqref{eq:RTE} and obtains
\begin{equation}
\mathbf I_{k+1}=\mathbf I_k+\int_{s_k}^{s_{k+1}}\mathbf F(s,\mathbf I(s)){\rm d}s\,,
\label{numerical_ftc}
\end{equation}
where the numerical approximation of a certain quantity at node $s_k$ is indicated by substituting the explicit dependence on $s$ with the subscript $k$, for instance
\begin{equation*}
\mathbf I_k \approx \mathbf I(s_k)\,.
\end{equation*}
Different approximations of the integral on the right-hand side of Equation~\eqref{numerical_ftc} yield different numerical methods. For the sake of generality,  this paper presents the different numerical methods in terms of the spatial coordinate $s$. All numerical schemes presented here can be straightforwardly formulated on geometrical or optical depth scale. The numerical analysis given in the following is not affected by this choice, unless otherwise specified. For instance, \citep[][hereafter referred to as Paper I]{janett2017} explained that the use of the optical depth scale usually mitigates fluctuations of the propagation matrix entries along the ray path, enforcing numerical stability.

Efficient integration schemes for Equation~\eqref{eq:RTE} or~\eqref{numerical_ftc} are of particular importance. The urgency of high-order well-behaved formal solvers has been soon recognized in the community and considerable efforts have been exercised in this direction. \citet{wittmann1974} and \citet{landi_deglinnocenti1976} first proposed high-order Runge-Kutta methods, which were then classified as very accurate at the expense of computational costs, because of the very small step size required \citep{landi_deglinnocenti1976,rees+al1989}. Thereafter, \citet{bellot_rubio+al1998} presented the fourth-order accurate (cubic) Hermitian method, showing its suitability as a formal solver. Later on, \citet{trujillo_bueno2003} argued that low-order schemes were inadequate to face the formal solution, showing the unsatisfying performance of DELO-linear \citep{rees+al1989} when applied to self-consistent non-LTE calculations and attempting to reach high-order convergence with DELOPAR. More recently, \citet{delacruz_rodriguez+piskunov2013} provided the fourth-order accurate DELO-B{\'e}zier methods and \citet{steiner2016} mentioned the possibility of using the high-order piecewise parabolic reconstruction when presenting their piecewise continuous method.

The many different high-order methods may produce some disorientation in the choice of a suitable formal solver. Therefore, continuing the analysis started by Paper I, this work attempts a clear characterization of the main high-order formal solvers. Section~\ref{sec:sec2} briefly presents the famous Runge-Kutta class, paying particular attention to the classical Runge-Kutta 4 method. Section~\ref{sec:sec3} introduces the linear multistep methods, focusing in particular on the Adams-Moulton family. Section~\ref{sec:sec4} is dedicated to Hermitian methods, where an insightful derivation of the cubic Hermitian method is presented. Section~\ref{sec:sec5} investigates the suitability of B{\'e}zier curves for the formal solution, highlighting an interesting connection to Hermitian methods. Finally, Section \ref{sec:sec6} provides remarks and conclusions, in an attempt to organize an effective hierarchy among formal solvers.
%
\section{Runge-Kutta methods}\label{sec:sec2}
Runge-Kutta methods form the best known class of one-step numerical schemes for ordinary differential equations. The formulas describing Runge-Kutta methods are abstracted away from the ideas of quadrature and collocation \citep{frank2008}. The basic idea of the Runge-Kutta methods is that there are many ways to evaluate the integral in Equation~\eqref{numerical_ftc}, and those methods all agree to low-order terms. The right combination of these gradually eliminates higher-order errors, increasing the order of accuracy. The general form of a $p$-stage (with $p\ge1$) Runge-Kutta method applied to Equation~\eqref{eq:RTE} reads \citep{frank2008}
\begin{equation*}
\mathbf I_{k+1}=\mathbf I_k+\sum_{i=1}^p b_i\mathbf k_i\,,
\end{equation*}
where $b_i$ are the weights, and the so-called stage values are given by
\begin{equation*}
\mathbf k_i=\Delta s_k\,\mathbf F\left(s_k+c_i\Delta s_k,\mathbf I_k+\sum_{j=1}^p a_{ij}\mathbf k_j\right)\,,\text{ for }i=1,\dots,p\,,
\end{equation*}
%
with $\Delta s_k=s_{k+1}-s_k$. The coefficients $[a_{ij}]$ form the Runge-Kutta matrix and the nodes $c_i$ lie in the interval $[0,1]$. A deeper look into this class of numerical methods is given, for instance, by \citet{deuflhard2002}.
\subsection{Runge-Kutta 4}
The best known high-order scheme of this class is probably the classical Runge-Kutta 4 method (RK4) and its application to the formal solution for polarized light was already proposed by \citet{landi_deglinnocenti1976}. The method is described by
\begin{equation}
\mathbf I_{k+1}=\mathbf I_k+\frac{1}{6}[\mathbf k_1 + 2\mathbf k_2 + 2\mathbf k_3+\mathbf k_4]\,,
\label{runge_kutta4}
\end{equation}
where the four stage values are given by
\begin{align*}
\mathbf k_1&=\Delta s_k\mathbf F(s_k,\mathbf I_k)\,,\\
\mathbf k_2&=\Delta s_k\mathbf F(s_k+\Delta s_k/2, \mathbf I_k+\mathbf k_1/2)\,,\\
\mathbf k_3&=\Delta s_k\mathbf F(s_k+\Delta s_k/2,\mathbf I_k+\mathbf k_2/2)\,,\\
\mathbf k_4&=\Delta s_k\mathbf F(s_{k+1},\mathbf I_k+\mathbf k_3)\,.
\end{align*}
The right-hand side of Equation~\eqref{runge_kutta4} does not contain the term $\mathbf I_{k+1}$ and RK4 is therefore classified as an explicit method.
\subsection{Order of accuracy}
RK4 is well known for its fourth-order accuracy, as indicated in Table~\ref{tab:convergence}, and its convergence analysis is easily found in the literature \citep[e.g.,][]{deuflhard2002,frank2008}. However, the quantities $\mathbf K$ and $\boldsymbol{\epsilon}$ at the intermediate point $s_k+\Delta s_k/2$ must be properly provided through interpolation. In order to maintain fourth-order accuracy, one needs an interpolation of degree $q\ge3$, i.e., at least cubic. A lower-order interpolation would decrease the order of accuracy: a parabolic interpolation results in a third-order accurate method and a linear interpolation provides a second-order accurate method. 

An alternative strategy for providing accurate absorption and emission quantities at the intermediate point is based on high-order interpolations of the atmospheric model parameters, such as the temperature, the microscopic and macroscopic velocities, and the strength and orientation of the magnetic field. The quantities $\mathbf K$ and $\boldsymbol{\epsilon}$ at $s_k+\Delta s_k/2$ are then evaluated through the interpolated thermodynamic parameters.
\subsection{Stability}
\begin{figure*}
\centering
\includegraphics[width=.8\textwidth]{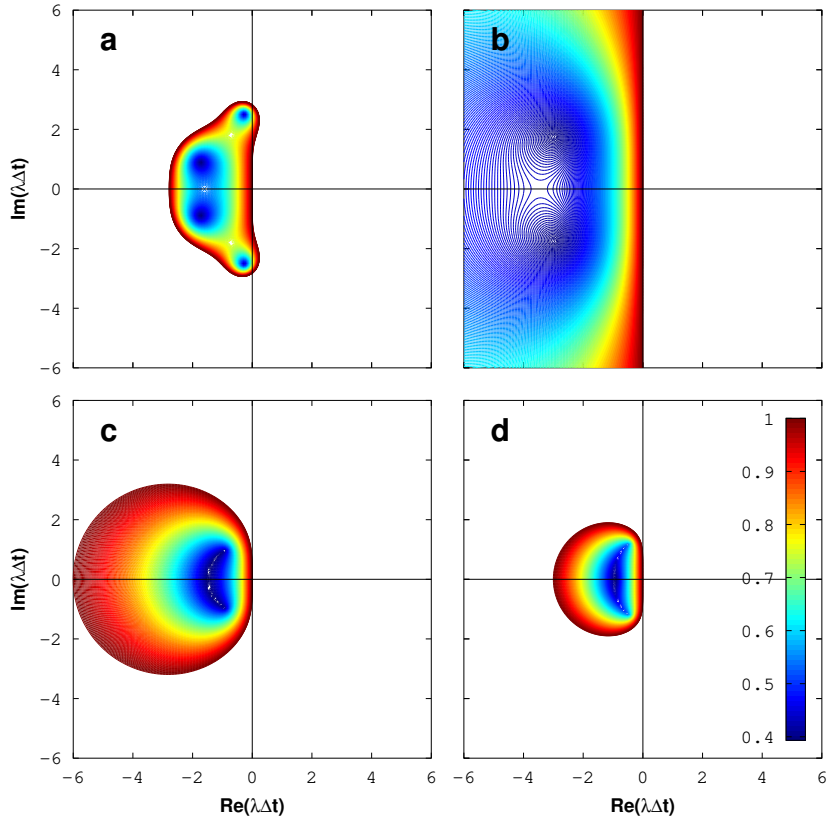}
  \caption{Stability regions for {\bf a)} the Runge-Kutta 4, {\bf b)} cubic Hermitian, {\bf c)} Adams-Moulton 3, and {\bf d)} Adams-Moulton 4 methods. The cubic Hermitian method shows A-stability, while all other methods have bounded stability regions.}
\label{fig:stability}
\end{figure*}

As explained in Paper I, the stability of a numerical method is often deduced through the simple autonomous scalar initial value problem (IVP) given by
\begin{equation}
\begin{aligned}
y'(t)&=\lambda y(t)\,,\\
y(0)&=y_0\,,
\label{IVP}
\end{aligned} 
\end{equation}
with $\lambda \in \mathbb{C}$. The solution $y(t)= y_0e^{\lambda t}$ converges to zero as $t \rightarrow \infty$ for $\operatorname{Re}(\lambda)< 0$. Defining $z = \lambda\Delta t$, where $\Delta t$ denotes the cell width, the RK4 method applied to the IVP \eqref{IVP} is recast into the form
\begin{equation*}
y_{k+1}=\phi_{\text{\tiny RK4}}(z)y_k\,,
\end{equation*}
where the stability function $\phi_{\text{\tiny RK4}}$ reads
\begin{equation*}
\phi_{\text{\tiny RK4}}(z)=1+z+\frac{z^2}{2}+\frac{z^3}{6}+\frac{z^4}{24}\,.
\end{equation*}
Stability is guaranteed by the condition $\Vert\phi_{\text{\tiny RK4}}(z)\Vert<1$. The stability region of RK4 presented in Figure~\ref{fig:stability}a is clearly bounded. This indicates that the method could suffer from magnification of numerical errors for large $z$ values. This problem is relevant for optically thick cells. In fact, the eigenvalues of the propagation operator $-\mathbf K$, which have real parts that are always negative, increase with the total absorption coefficient $\eta_I$ \citep{landi_deglinnocenti+landolfi2004}.

In order to face this problem, a hybrid technique can be used. This strategy applies an A-stable method, e.g., the trapezoidal method, for optical thick cells, making use of RK4 elsewhere. One correctly argues that this hybrid technique possesses the lowest order of accuracy among the used methods, i.e., second-order accuracy if the trapezoidal method is chosen. However, the strong attenuation induced by optically thick cells could reduce the error propagation. In fact, this hybrid method maintains fourth-order convergence, as clearly shown in Figure~\ref{convergence1} where it is labeled as Runge-Kutta 4.

Note that the assumption of a constant eigenvalue $\lambda$ in Equation~\eqref{IVP} is a limitation of this simplified stability analysis. In fact, variations of $\mathbf K$ along the integration path usually affect the stability region of the numerical method (see Paper I). Using an optical depth scale usually supports the assumption of a constant eigenvalue $\lambda$ in Equation~\eqref{IVP}.
\subsection{Computational cost}
As mentioned above, the RK4 method is explicit: it avoids the additional solution of the $4\times4$ implicit linear system, which is required by implicit methods. This fact could significantly reduce the total amount of computational time required. When applied to the polarized formal solution, RK4 has often been classified as computationally costly, because of the very small step size required \citep{rees+al1989}. In light of the previous stability analysis, one is led to believe that the requirement of small numerical cells is mainly due to the bounded stability region of the RK4 method and that the use of a hybrid strategy should overcome this problem.
%
%
\section{Linear multistep methods}\label{sec:sec3}
%
One-step methods compute the Stokes vector $\mathbf I_{k+1}$ solely on the basis of information about the preceding Stokes vector $\mathbf I_k$. In this sense, they have no memory, i.e., they forget all of the prior information that has been gained about the Stokes vector in previous steps. In contrast, multistep methods make use of the most recently found Stokes vectors $\mathbf I_{k-p+1},\dots,\mathbf I_{k}$ (with $p\ge1$) for the computation of $\mathbf I_{k+1}$. In the linear multistep class, a $p$-step method applied to Equation~\eqref{eq:RTE} can be always written in the form
\begin{equation}
\alpha_{k+1}\mathbf I_{k+1}=-\sum_{i=k-p+1}^{k}\alpha_i\mathbf I_i+\sum_{i=k-p+1}^{k+1}\beta_i\mathbf F_i\,,\text{ with }p\ge1\,.
\label{linear_multistep}
\end{equation}
If $\beta_{k+1}=0$, the numerical scheme is explicit, and if $\beta_{k+1}\neq0$, the scheme is implicit. This class of methods has been intensively investigated, in particular by the Swedish mathematician Germund Dahlquist (1925-2005). With his famous first and second barriers \citep{dahlquist1956,wanner2006}, he stated the lower stability of explicit methods in this class. Moreover, there are two main families of implicit linear multistep methods: Adams-Moulton methods and Backward Differentiation Formula methods. The latter are thought to increase stability, but are not as accurate as Adams-Moulton methods of the same order. Therefore, the implicit Adams-Moulton family is discussed only, following the adaptation to non-uniform spatial grids described by \citet{deuflhard2002}.
\subsection{Adams-Moulton methods}
The $p$-step Adams-Moulton's method approximates the integrand $\mathbf F$ in Equation~\eqref{numerical_ftc} by the $p$-order Lagrange polynomial
\begin{equation*}
\mathbf F(s,\mathbf I(s))\approx\sum_{i=k-p+1}^{k+1}\mathbf F_i\ell_i(s)\,\text{ for }s\in [s_k,s_{k+1}]\,,
\end{equation*}
which matches the numerical values $\mathbf F_i=-\mathbf K_i\mathbf I_i+\boldsymbol{\epsilon}_i$ at positions $s_i$. The Lagrange basis polynomials $\ell_i$ given by
\begin{equation*}
\ell_i(s)=\prod_{\substack{k-p+1\le m \le k+1\\m\neq i}}\frac{s-s_m}{s_i-s_m}\,,
\end{equation*}
satisfy the relation $\ell_i(s_j)=\delta_{ij}$, where the Kronecker delta $\delta_{ij}$ is defined by
\begin{equation*}
\delta_{ij} = 
\begin{cases}
1 & \text{ if } i = j \,, \\ 
0 & \text{ if } i \ne j\,.
\end{cases}
\end{equation*}
The integral in Equation~\eqref{numerical_ftc} can then be solved by parts, yielding, after some algebra, a linear system of the form of Equation~\eqref{linear_multistep}. The presence of the term $\mathbf F_{k+1}$ in the Lagrange interpolation provides $\beta_{k+1}\neq0$, indicating the method to be implicit. By contrast, Adams-Bashford methods define the Lagrange interpolation through $\mathbf F_{k-p},\dots,\mathbf F_k$, providing a linear system with $\beta_{k+1}=0$, which is therefore explicit.

At this point, one has to note the critical difference between the Adams-Moulton family and the DELO family discussed in Paper I. In the former, the Lagrange polynomial approximates the integrand $\mathbf F$, while in the latter the interpolation is applied to the effective source function. Nonetheless, the two different strategies share similar convergence properties, because the local truncation error depends on the interpolation degree in both cases.

Although outside the assumption $p\ge1$, it is habit to include the $p=0$ case in the Adams-Moulton family. In this instance, the integrand in Equation~\eqref{numerical_ftc} is approximated as $\mathbf F\approx\mathbf F_{k+1}$. Doing so, one obtains the common first-order accurate backward (or implicit) Euler method \citep[e.g.,][]{deuflhard2002}.

Now, if the first-order Lagrange interpolant is used to approximate the integrand $\mathbf F$ in Equation~\eqref{numerical_ftc}, one obtains the famous second-order accurate trapezoidal method as assessed in Paper I. Note that both the backward Euler method and the trapezoidal method are one-step methods, which also belong to the Runge-Kutta class (see Section~\ref{sec:sec2}).

If a parabolic Lagrange interpolation is performed through $\mathbf F_{k-1}$, $\mathbf F_k$, and $\mathbf F_{k+1}$, one obtains the implicit linear system given by
\begin{equation}
\mathbf{\Phi}_{k+1}\mathbf{I}_{k+1}=\mathbf{\Phi}_k\mathbf{I}_k+\mathbf{\Phi}_{k-1}\mathbf{I}_{k-1}
+\mathbf{\Psi}_{k+1}+\mathbf{\Psi}_k+\mathbf{\Psi}_{k-1}\,,
\label{adams_moulton_3}
\end{equation}
and the coefficients $\mathbf{\Phi}_{k-1}$, $\mathbf{\Phi}_k$, $\mathbf{\Phi}_{k+1}$, $\mathbf{\Psi}_{k-1}$, $\mathbf{\Psi}_k$, and $\mathbf{\Psi}_{k+1}$ are provided in Appendix~\ref{appendix:B}. The two-step numerical scheme described by Equation~\eqref{adams_moulton_3} is called Adams-Moulton 3 method.

A cubic Lagrange interpolation through $\mathbf F_{k-2}$, $\mathbf F_{k-1}$, $\mathbf F_k$, and $\mathbf F_{k+1}$, provides the following implicit linear system
\begin{equation}
\begin{split}
\mathbf{\Phi}_{k+1}\mathbf{I}_{k+1}&=\mathbf{\Phi}_k\mathbf{I}_k+\mathbf{\Phi}_{k-1}\mathbf{I}_{k-1}+\mathbf{\Phi}_{k-2}\mathbf{I}_{k-2}\\
&	+\mathbf{\Psi}_{k+1}+\mathbf{\Psi}_k+\mathbf{\Psi}_{k-1}+\mathbf{\Psi}_{k-2}\,,
\label{adams_moulton_4}
\end{split}
\end{equation}
and the coefficients $\mathbf{\Phi}_{k-2}$, $\mathbf{\Phi}_{k-1}$, $\mathbf{\Phi}_k$, $\mathbf{\Phi}_{k+1}$, $\mathbf{\Psi}_{k-2}$,$\mathbf{\Psi}_{k-1}$, $\mathbf{\Psi}_k$, and $\mathbf{\Psi}_{k+1}$ are provided in Appendix~\ref{appendix:B}. The three-step numerical scheme described by Equation~\eqref{adams_moulton_4} is called Adams-Moulton 4 method.

This family of formal solvers can be further expanded by just increasing the interpolation degree of the integrand $\mathbf F$. However, the complexity of the numerical methods would increase and the expressions for the $\mathbf{\Phi}$ and $\mathbf{\Psi}$ coefficients would become gradually more cumbersome.
\subsection{Order of accuracy}
\begin{figure*}
\centering
\includegraphics[width=.8\textwidth]{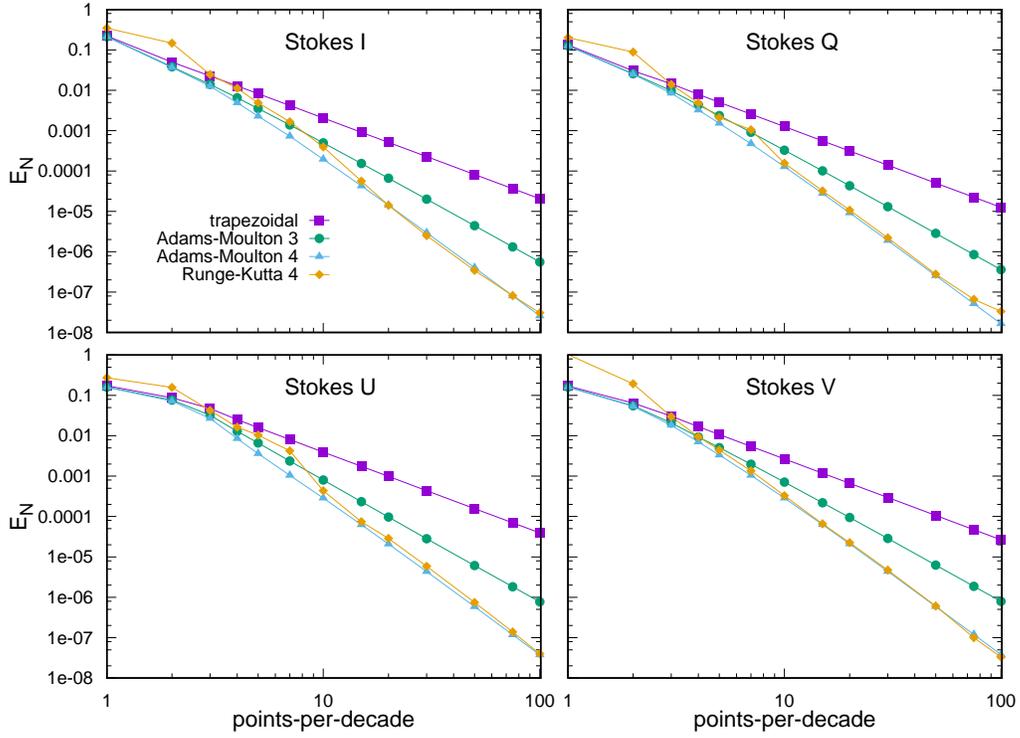}
  \caption{log-log representation of the global error for the Stokes vector components $I,Q,U$ and $V$ as functions of the number of points-per-decade of the continuum optical depth for the trapezoidal, Adams-Moulton 3 and 4, and RK4 methods. The atmospheric model and the spectral line parameters are identical to those described in Appendix C of Paper I and the error is calculated as described in Appendix D of Paper I. Note that while the absolute value and the pre-asymptotic behavior depend on the specific atmospheric model, the order of accuracy, i.e., the slope of the curves in the asymptotic regime, is not.}
\label{convergence1}
\end{figure*}

The local truncation error of the Adams-Moulton methods is due to the fact that the integrand $\mathbf F$ in Equation~\eqref{numerical_ftc} is approximated by a polynomial. A Lagrange polynomial of degree $p$ is known to be $(p+1)$th-order accurate. The resulting local truncation error satisfies
\begin{equation*}
L^{\text{\tiny A-M}}[p] \approx O(\Delta s^{p+2})\,,
\end{equation*}
indicating a $p$-step Adams-Moulton method as $(p+1)$-order accurate (see Paper I). Accordingly, the Adams-Moulton 3 method described by Equation~\eqref{adams_moulton_3} is third-order accurate, whereas the Adams-Moulton 4 method described by Equation~\eqref{adams_moulton_4} is fourth-order accurate, as summarized in Table~\ref{tab:convergence} and a numerical confirmation is given by Figure~\ref{convergence1}.
\subsection{Stability}
The simple stability analysis performed above for one-step methods cannot be applied to this class, because of the multiterm contribution. Therefore, linear multistep methods require a more complex derivation of the stability region \citep{frank2008}.

The numerical scheme given by Equation~\eqref{linear_multistep} applied to the IVP \eqref{IVP} gives
\begin{equation*}
\sum_{i=k-p+1}^{k+1}\alpha_i y_i=\lambda\Delta t\sum_{i=k-p+1}^{k+1}\beta_i y_i\,,
\end{equation*}
and, defining $z = \lambda\Delta t$, one gets
\begin{equation*}
\sum_{i=k-p+1}^{k+1} \left( \alpha_i - z\,\beta_i \right) y_i = 0\,.
\end{equation*}
For any $z$, this is a linear difference equation with the characteristic polynomial
\begin{equation}
\sum_{i=k-p+1}^{k+1}\left(\alpha_i - z\,\beta_i\right)\zeta^i = 0 = \rho(\zeta)-z\,\sigma(\zeta).
\label{polynomial}
\end{equation}
The stability region of a linear multistep method is the set of complex values $z$ for which all roots $\zeta$ of the polynomial Equation~\eqref{polynomial} lie on the unit disk, i.e. $|\zeta|\le1$, and those with modulus one are simple. On the boundary of the stability region, precisely one root has modulus one. Therefore, an explicit representation for the boundary of the stability region is given by
\begin{equation*}
\delta S= \left\{z=\frac{\rho(e^{i\theta})}{\sigma(e^{i\theta})}\right\}\,,\text{ for }\theta\in[-\pi,\pi].
\end{equation*}
The stability regions of the Adams-Moulton 3 and Adams-Moulton 4 methods are clearly bounded, as shown by Figures~\ref{fig:stability}c and~\ref{fig:stability}d. As in the case of the RK4 method, stiffness could appear in optically thick cells, imposing a reduction of the cell width or a switch to A-stable methods to maintain convergence. Therefore, stability constraints are clearly a disadvantage when using high-order Adams-Moulton methods. In this sense, the second Dahlquist barrier clarifies the situation, stating that an A-stable linear multistep method has an order of accuracy $p\le2$.

Although the use of optical depth usually supports the assumption of a constant eigenvalue $\lambda$ in Equation~\eqref{IVP}, this assumption limits the validity of the stability analysis. An additional limitation is given by the fact that the stability analysis of linear multistep methods assumes a homogeneous discrete grid. Strongly variable meshes, such as logarithmically spaced grids, could therefore alter the stability conditions for linear multistep schemes.
\subsection{Computational cost}
As mentioned above, the Adams-Moulton methods are implicit and require the solution of a $4\times4$ implicit linear system. The similarity to the DELO methods suggests a similar computational cost.
\begin{figure*}
\centering
\includegraphics[width=.8\textwidth]{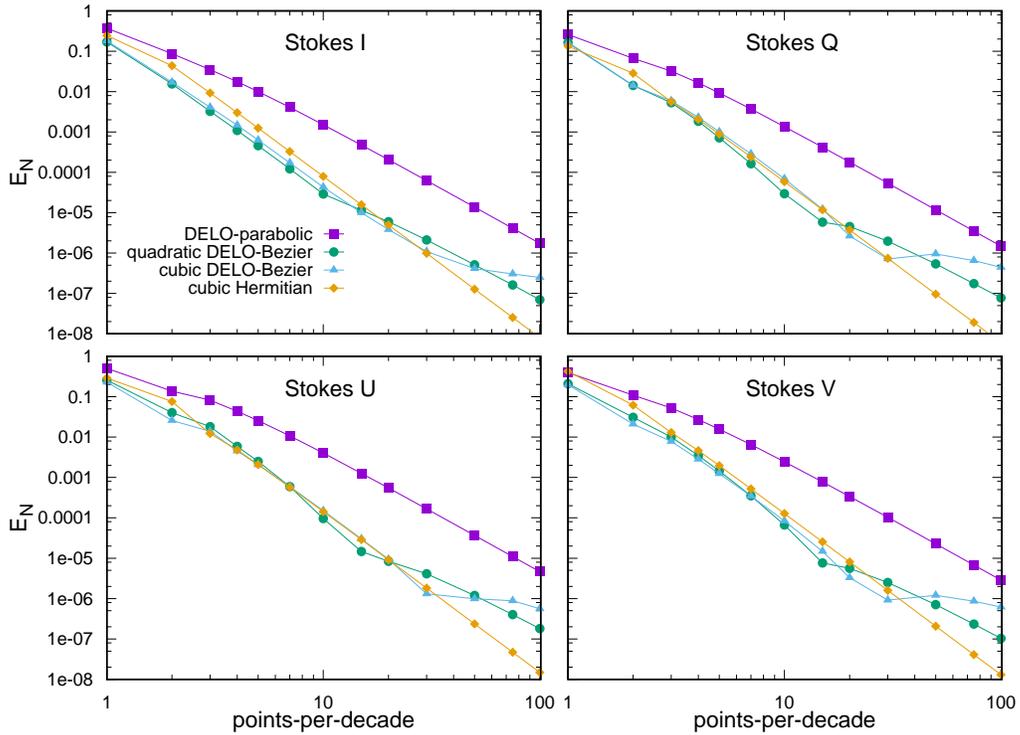}
  \caption{Same as Figure~\ref{convergence1}, but for different methods, namely: the DELO-parabolic, quadratic and cubic DELO-B{\'e}zier, and cubic Hermitian methods.}
\label{convergence2}
\end{figure*}
\section{Hermitian methods}\label{sec:sec4}
Adams-Moulton methods approximate the integrand $\mathbf F$ in terms of Lagrange polynomials. However, the literature provides different interpolation strategies and the set of suitable interpolants proposed by \citet{auer2003} for the scalar formal solution includes Hermite polynomials. 

Given a set of points $\{x_i\}$ $(i=1,\dots,n)$, the Hermite interpolation $H$ does not match only a set of function values $\{y_i\}$, but also its derivatives, i.e.,
\begin{equation*}
H^{(k)}(x_i)=y^{(k)}_i\,,
\end{equation*}
for $k=0,\dots,m_i-1$ and $i=1,\dots,n$. The minimal degree $q$ of the Hermite polynomial, which can satisfy the conditions given above, is given by
\begin{equation*}
q=\sum_{i=1}^{n}m_i\,.
\end{equation*}
This section focuses on the use of the cubic Hermitian interpolation to approximate the integrand in Equation~\eqref{numerical_ftc}, where both grid values and first derivatives of $\mathbf F$ are specified at the nodes $s_k$ and $s_{k+1}$. If, in addition, the second derivatives of $\mathbf F$ are available, the quintic Hermitian interpolation can be used. However, the algorithm complexity would increase, raising some doubts on its suitability for the polarized radiative transfer problem.
\subsection{Cubic Hermitian method}
Here, the cubic Hermite interpolation is chosen to approximate $\mathbf F$ in Equation~\eqref{numerical_ftc}. For notational simplicity one defines the normalized variable $t \in [0,1]$ as
\begin{equation*}
t=\frac{s-s_k}{\Delta s_k}\,,\text{ for } s \in [s_k,s_{k+1}]\,.
\end{equation*}
The cubic Hermite interpolation, approximating $\mathbf F$ inside the interval $[0,1]$, reads
\begin{equation}
\begin{split}
\mathbf F(t)&\approx\mathbf F_k\cdot(1-3t^2+2t^3) + \mathbf F_k'\cdot \Delta s_k(t-2t^2+t^3)\\
&+ \mathbf F_{k+1}\cdot (3t^2-2t^3) + \mathbf F_{k+1}'\cdot \Delta s_k(-t^2+t^3)\,.
\label{F_hermite_cubic}
\end{split}
\end{equation}
In addition to the grid values $\mathbf F_k$ and $\mathbf F_{k+1}$, the first derivatives $\mathbf F_k'$ and $\mathbf F_{k+1}'$ are also specified and Equation~\eqref{F_hermite_cubic} provides the unique third-degree polynomial that matches both node values and node first derivatives at $s_k$ and $s_{k+1}$. Moreover, the first derivative of $\mathbf F$ satisfies
\begin{align*}
\mathbf F'(s)&=-\mathbf K'(s)\mathbf I(s)-\mathbf K(s)\mathbf I'(s)+\boldsymbol{\epsilon}'(s)\\
&= \left[\mathbf K(s)\mathbf K(s)-\mathbf K'(s)\right]\mathbf I(s)-\mathbf K(s)\boldsymbol{\epsilon}(s)+\boldsymbol{\epsilon}'(s)\,,
\end{align*}
where Equation~\eqref{eq:RTE} is used to replace the Stokes vector first derivative $\mathbf I'$. Inserting numerical approximations, the first derivatives $\mathbf F_k'$ and $\mathbf F_{k+1}'$ can be written as
\begin{align*}
\mathbf F'_k & = \left[\mathbf K_k\mathbf K_k-\mathbf K'_k\right]\mathbf I_k-\mathbf K_k\boldsymbol{\epsilon}_k+\boldsymbol{\epsilon}'_k\,,\\
\mathbf F'_{k+1} &= \left[\mathbf K_{k+1}\mathbf K_{k+1}-\mathbf K'_{k+1}\right]\mathbf I_{k+1} -\mathbf K_{k+1}\boldsymbol{\epsilon}_{k+1} +\boldsymbol{\epsilon}'_{k+1}\,.
\end{align*}
Replacing the integrand $\mathbf F$ in Equation~\eqref{numerical_ftc} with the cubic Hermite interpolant given by Equation~\eqref{F_hermite_cubic}, one evaluates the integral by parts. Making use of the previous identities for $\mathbf F'_k$ and $\mathbf F'_{k+1}$ and performing some algebra, one recovers the following implicit linear system
\begin{equation}
\mathbf{\Phi}_{k+1}\mathbf I_{k+1}=\mathbf{\Phi}_k\mathbf I_k+\mathbf{\Psi}_{k+1}+\mathbf{\Psi}_k\,,
\label{recursive_hermite}
\end{equation}
where
\begin{align*}
\mathbf{\Phi}_k&=\mathbf{1}-\frac{\Delta s_k}{2}\mathbf K_k+\frac{\Delta s_k^2}{12}\left[\mathbf K_k\mathbf K_k-\mathbf K'_k\right]\,,\\
\mathbf{\Phi}_{k+1}&=\mathbf{1}+\frac{\Delta s_k}{2}\mathbf K_{k+1}+\frac{\Delta s_k^2}{12}\left[\mathbf K_{k+1}\mathbf K_{k+1}-\mathbf K'_{k+1}\right]\,,\\
\mathbf{\Psi}_k&=\frac{\Delta s_k}{2}\boldsymbol{\epsilon}_k+\frac{\Delta s_k^2}{12}\left[\boldsymbol{\epsilon}'_k-\mathbf K_k\boldsymbol{\epsilon}_k\right]\,,\\
\mathbf{\Psi}_{k+1}&=\frac{\Delta s_k}{2}\boldsymbol{\epsilon}_{k+1}-\frac{\Delta s_k^2}{12}\left[\boldsymbol{\epsilon}'_{k+1}-\mathbf K_{k+1}\boldsymbol{\epsilon}_{k+1}\right]\,.
\end{align*}
The one-step numerical method described by Equation~\eqref{recursive_hermite} corresponds exactly to the one proposed by \citet{bellot_rubio+al1998}, but here it is derived through a different strategy. Moreover, the first derivatives $\mathbf K'$ and $\boldsymbol{\epsilon}'$ are usually not provided by the problem and must be numerically approximated. The accuracy of the numerical derivatives could affect the order of accuracy of the entire method and \citet{bellot_rubio+al1998} first opted for an expensive procedure based on cubic spline interpolation. When considering a physical quantity $u$, they also mentioned the possibility of calculating the numerical first derivative $u_k'$ at the node $s_k$, assuming a parabolic dependence along $s_{k-1}$, $s_k$, and $s_{k+1}$. The explicit formula adapted to a non-uniform spatial grid reads
\begin{equation}
u_k'=w_{k-1}u_{k-1}+w_k u_k+w_{k+1}u_{k+1}\,,
\label{num_der}
\end{equation}
where
\begin{align*}
w_{k-1}&=\frac{1}{\Delta s_{k+1}+\Delta s_k}-\frac{1}{\Delta s_k}\,,\\
w_k&=\frac{1}{\Delta s_k}-\frac{1}{\Delta s_{k+1}}\,,\\
w_{k+1}&=\frac{1}{\Delta s_{k+1}}-\frac{1}{\Delta s_{k+1}+\Delta s_k}\,,
\end{align*}
which is a second-order accurate approximation for the first derivatives. \citet{fritsch1984} proposed an alternative formula to recover second-order accurate first derivatives for producing monotone piecewise cubic Hermite interpolants.
\begin{table}
\caption{Order of accuracy for different high-order methods}
\setlength{\tabcolsep}{5pt}\renewcommand{\arraystretch}{1.5}
\centering
\begin{tabular}{|l|c|c|}\hline
\emph{Formal solver}	& \emph{Order of accuracy}\\
\hline 
Runge-Kutta 4		& 4 \\
Adams-Moulton 3		& 3 \\
Adams-Moulton 4		& 4 \\
Cubic Hermitian		& 4 \\
DELO-parabolic		& 3 \\
Quadratic DELO-B\'ezier	& 4 \\
Cubic DELO-B\'ezier	& 4 \\\hline 
\end{tabular}
\label{tab:convergence}
\end{table}
\subsection{Order of accuracy}
The local truncation error is due to the fact that the integrand $\mathbf F$ is approximated by a cubic Hermite polynomial. One can show that the cubic Hermite interpolant is fourth-order accurate if the derivatives are at least third-order, third-order if the  derivatives are second-order, and so on \citep{dougherty1989}. Therefore, assuming derivatives are at least third-order accurate, the global error scales as $O(\Delta s^4)$, indicating the cubic Hermitian method as fourth-order accurate (see Table~\ref{tab:convergence}). In confirmation of this, \citet{bellot_rubio+al1998} perform an alternative convergence analysis, providing the same order of accuracy. The local truncation error analysis based on Taylor expansion proposed in Appendix~\ref{appendix:A} reveals that second-order accurate numerical derivatives, such as, for instance, the one given by Equation \eqref{num_der}, are already sufficient to maintain the fourth-order accuracy of the cubic Hermitian method, as confirmed by Figure~\ref{convergence2}. However, \cite{delacruz_rodriguez+piskunov2013} indicate the cubic Hermitian method as third-order accurate. It can be surmised that this is due to the first-order accurate derivatives used in the method there, which are not sufficient to maintain fourth-order accuracy.
\subsection{Stability}
The stability function of the cubic Hermitian method is easily deduced through the IVP \eqref{IVP} and reads
\begin{equation*}
\phi_{\text{\tiny H}}(z)=\frac{1+z/2+z^2/12}{1-z/2+z^2/12}\,,
\end{equation*}
with $z = \lambda\Delta t$. Stability is then given by the condition $\Vert\phi_{\text{\tiny H}}(z)\Vert<1$. As displayed in Figure~\ref{fig:stability}b, the stability region contains the whole left-hand side of the complex plane, indicating the cubic Hermitian method as A-stable. Paper I argues that this is an important feature to avoid numerical instability in the formal solution.

Once more, the assumption of a constant eigenvalue $\lambda$ in Equation~\eqref{IVP} is a limitation for the stability analysis because variations of $\mathbf K$ along the integration path affect the stability region of the cubic Hermitian method. The use of the optical depth scale usually supports the assumption of a constant eigenvalue $\lambda$ in Equation~\eqref{IVP}.
\subsection{Computational cost}
The cubic Hermitian method is implicit and it requires the solution of the $4\times4$ linear system given by Equation~\eqref{recursive_hermite}. The additional matrix-by-matrix multiplications and the calculation of numerical derivatives increase the computational effort. However, \citet{bellot_rubio+al1998} suggest that, when a certain accuracy is required, the high accuracy of the cubic Hermitian method allows one to use coarser spatial grids, reducing the total computational cost of the problem.
\section{B{\'e}zier methods}\label{sec:sec5}
In addition to the Hermitian interpolation, \citet{auer2003} mentioned the possibility of using B{\'e}zier curves in the formal solution, aiming to suppress spurious extrema.  These interpolations, named after Pierre B{\'e}zier (1910-1999), make use of the so-called control points (or weights).

\begin{figure}
\plotone{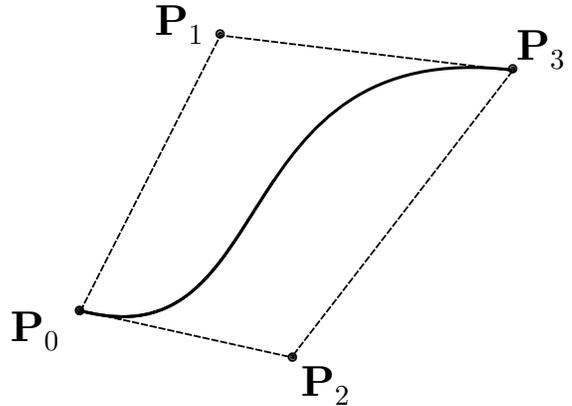}
  \caption{The solid curve represents a cubic B\'ezier curve with the start point $\mathbf P_0$, the two weights $\mathbf P_1$ and $\mathbf P_2$, and the end point $\mathbf P_3$. The dashed lines delimit the convex hull of the control points.}
\label{fig:bezier}
\end{figure}

A B{\'e}zier curve of degree $q$ applied to the integrand $\mathbf F$ in Equation~\eqref{numerical_ftc} can be defined as
\begin{equation*}
\mathbf B_q(t)=\sum_{i=0}^{q}\mathbf P_i B_{i,q}(t)\,,
\end{equation*}
where $t \in[0,1]$, $\mathbf P_i$ are the control points, and the Bernstein polynomials $B_{i,q}$ are given by
\begin{equation*}
B_{i,q}(t)=\binom{q}{i}\cdot t^i\left(1-t\right)^{q-i}\,.
\end{equation*}
The first and the last control points define the start and end points of the B{\'e}zier curve, i.e.,
\begin{equation*}
\mathbf P_0 = \mathbf F_k\,,\text{ and }\mathbf P_q = \mathbf F_{k+1}\,.
\end{equation*}
All the remaining points, conventionally called weights, are usually used to shape the curve. When aiming to increase accuracy, B{\'e}zier interpolants are usually forced to be identical to Hermite interpolants by a proper tuning of the weights. Moreover, a B{\'e}zier curve always lies in the convex hull of the control points, i.e., in the smallest set that contains the line segment joining every pair of control points. This property can be used to avoid the creation of new extrema by adjusting the weights and it is suitable to prevent spurious behavior near rapid changes in the absorption and emission coefficients, preserving monotonicity in the interpolation. An illustrative example of a cubic B\'ezier curve is given by Figure~\ref{fig:bezier}.

B{\'e}zier methods are therefore based on interpolations that avoid overshooting when treating intermittent quantities and correspond to Hermitian interpolations when considering smooth ones. This strategy is very similar to the one proposed by \citet{delacruz_rodriguez+piskunov2013} for DELO-B{\'e}zier methods, where the B{\'e}zier interpolation is applied to the effective source function instead of to $\mathbf F$. The two different strategies share similar convergence properties, because the local truncation error originates from the polynomial approximation in both cases. However, the strategy presented in this section maintains a simpler form, avoiding the use of exponential functions and the problematic division of vanishingly small quantities.

If the linear B{\'e}zier curve, which is just a straight-line, is used to approximate $\mathbf F$ inside the interval $[0,1]$, one simply obtains the trapezoidal method. However, quadratic and cubic B{\'e}zier interpolants deserve a deeper investigation.
\subsection{Quadratic B{\'e}zier method}
If the quadratic B{\'e}zier curve is used to approximate $\mathbf F$ inside the interval $[0,1]$, one gets
\begin{equation}
\mathbf F(t)\approx \mathbf F_k\cdot (1-t)^2 + \mathbf C\cdot 2 t(1-t) + \mathbf F_{k+1}\cdot t^2\,.
\label{F_bezier_quadratic}
\end{equation}
While the grid values $\mathbf F_k$ and $\mathbf F_{k+1}$ determine the start and the end points of the curve, the presence of the control point $\mathbf C$ allows one to shape the B{\'e}zier curve inside the interval. \citet{auer2003} proposed two different expressions for the control point $\mathbf C$, such that the quadratic B{\'e}zier curve corresponds to a quadratic Hermite polynomial, namely,
\begin{align*}
\mathbf C^{(A)} &= \mathbf F_k+\frac{\Delta s_k}{2}\mathbf F_k'\,,\\
\mathbf C^{(B)} &= \mathbf F_{k+1}-\frac{\Delta s_k}{2}\mathbf F_{k+1}'\,,
\end{align*}
intending to maximize the accuracy of the interpolation. In fact, \citet{auer2003} points out that from the standpoints of continuity and accuracy, Hermite interpolation is preferred. Moreover, \citet{delacruz_rodriguez+piskunov2013} suggest that if both $\mathbf C^{(A)}$ and $\mathbf C^{(B)}$ can be computed, it is desirable to take the mean, i.e.,
\begin{equation}
\mathbf C=\frac{\mathbf C^{(A)}+\mathbf C^{(B)}}{2}\,,
\label{control_point}
\end{equation}
recovering a more symmetric interpolation. Replacing the integrand $\mathbf F$ in Equation~\eqref{numerical_ftc} with the quadratic B{\'e}zier interpolant given by Equation~\eqref{F_bezier_quadratic} and inserting the symmetric control point given by Equation~\eqref{control_point}, one evaluates the integral by parts. Making use of the identities for $\mathbf F'_k$ and $\mathbf F'_{k+1}$ given in the previous section and performing some algebra, one recovers the implicit linear system described by Equation~\eqref{recursive_hermite}. Therefore, the obtained method corresponds exactly to the cubic Hermitian method, sharing its order of accuracy, stability, and computational cost properties.

This result intuitively explains the fourth-order accuracy obtained by the quadratic DELO-B{\'e}zier method described by \citet{delacruz_rodriguez+piskunov2013}, for which the complexity of the method prevents an analytical prediction of the order of accuracy.
\subsection{Cubic B{\'e}zier method}
If the cubic B{\'e}zier curve is used to approximate $\mathbf F$ inside the interval $[0,1]$, one gets
{\small
\begin{equation}
\mathbf F(t)\approx\mathbf F_k\cdot (1-t)^3 + \mathbf{\tilde C}_k\cdot 3t(1-t)^2 + \mathbf{\tilde C}_{k+1}\cdot 3t^2(1-t) + \mathbf F_{k+1}\cdot t^3\,.
\label{F_bezier_cubic}
\end{equation}}\noindent
The cubic B{\'e}zier curve is forced to be identical to the cubic Hermite polynomial given by Equation~\eqref{F_hermite_cubic} by adopting the following control points,
\begin{align*}
\mathbf{\tilde C}_k &= \mathbf F_k+\frac{\Delta s_k}{3}\mathbf F_k'\,,\\
\mathbf{\tilde C}_{k+1} &= \mathbf F_{k+1}-\frac{\Delta s_k}{3}\mathbf F_{k+1}'\,.
\end{align*}
Therefore, if the integrand $\mathbf F$ in Equation~\eqref{numerical_ftc} is replaced by the cubic B{\'e}zier curve given by Equation~\eqref{F_bezier_cubic} with the control points specified above, the resulting implicit linear system corresponds, once again, to the one given by Equation~\eqref{recursive_hermite}.
\section{Conclusions}\label{sec:sec6}
This paper exposes and compares different high-order candidate methods for the numerical evaluation of the radiative transfer equation for polarized light. The performed analysis highlights the advantages and the weaknesses of the considered numerical schemes, allowing some objective assessments.

The explicit RK4 method is fourth-order accurate. In this scheme, one has to provide the propagation matrix and the emission vector at an intermediate node in the computational cell. In order to maintain high-order accuracy, these quantities must be obtained through high-order interpolations. RK4 suffers from instability when treating optically thick cells. This fact could impose a reduction of the cell width, because instabilities either lead to a deterioration of accuracy or prevent convergence. This problem is circumvented by a hybrid technique that switches to the A-stable trapezoidal method when stiffness appears. RK4 remains competitive through the force of its reduced computational cost, because it avoids the solution of the $4\times4$ linear system.

The multistep Adams-Moulton strategy reaches high-order accuracy. In this work, the third-order and fourth-order Adams-Moulton methods are exposed. Both methods share similar accuracy and instability issues with the RK4 method, but they are computationally more expensive. Moreover, a $p$-step method requires the $p$ most recently found Stokes vectors. No clear improvement is brought with respect to RK4: therefore this class of methods is not recommended for the high-order numerical evaluation of Equation~\eqref{eq:RTE}.

The cubic Hermitian method, first applied to the polarized radiative transfer by \citet{bellot_rubio+al1998}, seems to be a good candidate, because of its fourth-order accuracy and A-stability. Moreover, the first derivatives $\mathbf K'$ and $\boldsymbol{\epsilon}'$ must be provided, and this is usually done through interpolation: here, a parabolic interpolant is sufficient to maintain fourth-order accuracy. A possible weakness of this method is the computational cost: the matrix-by-matrix multiplications required and the calculation of numerical derivatives described above could significantly increase the total computational effort.

Regarding B{\'e}zier methods, some considerations are necessary. First of all, the high-order convergence of B{\'e}zier methods is guaranteed by forcing the B{\'e}zier interpolants to be identical to the corresponding degree Hermite interpolants when approximating $\mathbf F$ in Equation~\eqref{numerical_ftc} and providing at least second-order accurate derivatives. Second, the usefulness of B{\'e}zier polynomials lies in their ability to remove spurious extrema. This feature is fundamental when reconstructing positive physical quantities from discrete values \citep[e.g.,][]{auer2003,ibgui2013}, but its effective benefit when approximating the integrand $\mathbf F$ in Equation~\eqref{numerical_ftc} has not yet been proven. Third, the detection of local extrema requires conditional \emph{if...else} statements, which burden the algorithm. In view of the absence of explicit supporting results, the use of B{\'e}zier polynomials in the numerical integration of Equation~\eqref{eq:RTE} is not supported.

The DELO family also provides different high-order formal solvers. Provided the same considerations previously made for B{\'e}zier methods, quadratic and cubic DELO-B{\'e}zier methods usually perform as fourth-order accurate methods (see Figure~\ref{convergence2} and Table~\ref{tab:convergence}). Paper I also explains that the DELO strategy is thought to remove stiffness from the problem. However, a deeper stability comparison with the A-stable cubic Hermitian method remains to be explored. Moreover, DELO methods always require the evaluation of coefficients, which include exponential terms, making the algorithm more involved \citep[e.g., the problematic division of vanishingly small quantities described by][]{delacruz_rodriguez+piskunov2013}.

The effective performance of the numerical methods when dealing with realistic atmospheric models remains to be explored.
\acknowledgments 
The financial support by the Swiss National Science Foundation (SNSF) through grant ID 200021\_159206/1 is gratefully acknowledged. Special thanks are extended to F. Calvo and A. Paganini for particularly enriching discussions.
%
\appendix
\section{Adams-Moulton coefficients}\label{appendix:B}
The coefficients of the Adams-Moulton 3 method, Equation~\eqref{adams_moulton_3}, are given by
\begin{align*}
\mathbf{\Phi}_{k-1}&=-\Phi_{k-1}\mathbf K_{k-1}\,,\\
\mathbf{\Phi}_k&=\mathbf 1-\Phi_k\mathbf K_k\,,\\
\mathbf{\Phi}_{k+1}&=\mathbf 1+\Phi_{k+1}\mathbf K_{k+1}\,,\\
\mathbf{\Psi}_{k-1}&=\Phi_{k-1}\boldsymbol{\epsilon}_{k-1}\,,\\
\mathbf{\Psi}_k&=\Phi_k\boldsymbol{\epsilon}_k\,,\\
\mathbf{\Psi}_{k+1}&=\Phi_{k+1}\boldsymbol{\epsilon}_{k+1}\,,
\end{align*}
with
\begin{align*}
\Phi_{k-1}&=-\frac{\Delta s_k^3}{6\Delta s_{k-1}(\Delta s_{k-1}+\Delta s_k)}\,,\\
\Phi_k&=-\frac{\Delta s_k(3\Delta s_{k-1}+\Delta s_k)}{6\Delta s_{k-1}}\,,\\
\Phi_{k+1}&=\frac{\Delta s_k(3\Delta s_{k-1}+2\Delta s_k)}{6(\Delta s_{k-1}+\Delta s_k)}\,.
\end{align*}
The coefficients of the Adams-Moulton 4 method, Equation~\eqref{adams_moulton_4}, are given by
\begin{align*}
\mathbf{\Phi}_{k-2}&=-\Phi_{k-2}\mathbf K_{k-2}\,,\\
\mathbf{\Phi}_{k-1}&=-\Phi_{k-1}\mathbf K_{k-1}\,,\\
\mathbf{\Phi}_k&=\mathbf 1-\Phi_k\mathbf K_k\,,\\
\mathbf{\Phi}_{k+1}&=\mathbf 1+\Phi_{k+1}\mathbf K_{k+1}\,,\\
\mathbf{\Psi}_{k-2}&=\Phi_{k-2}\boldsymbol{\epsilon}_{k-2}\,,\\
\mathbf{\Psi}_{k-1}&=\Phi_{k-1}\boldsymbol{\epsilon}_{k-1}\,,\\
\mathbf{\Psi}_k&=\Phi_k\boldsymbol{\epsilon}_k\,,\\
\mathbf{\Psi}_{k+1}&=\Phi_{k+1}\boldsymbol{\epsilon}_{k+1}\,,
\end{align*}
with
{\small
\begin{align*}
\Phi_{k-2}&=\frac{\Delta s_k^3(\Delta s_k+2\Delta s_{k-1})}{12\Delta s_{k-2}(\Delta s_{k-2}+\Delta s_{k-1})(\Delta s_{k-2}+\Delta s_{k-1}+\Delta s_k)}\,,\\
\Phi_{k-1}&=-\frac{\Delta s_k^3(\Delta s_k+2\Delta s_{k-1}+2\Delta s_{k-2})}{12\Delta s_{k-2}\Delta s_{k-1}(\Delta s_{k-1}+\Delta s_k)}\,,\\
\begin{split}
\Phi_k &=\frac{\Delta s_k}{12\Delta s_{k-1}(\Delta s_{k-2}+\Delta s_{k-1})}\Bigl[(\Delta s_k+2\Delta s_{k-1}+\Delta s_{k-2})^2\\
&+\Delta s_{k-1}^2+(\Delta s_{k-1}+\Delta s_{k-2})^2-2\Delta s_{k-2}^2\Bigr]\,,
\end{split}\\
\begin{split}
\Phi_{k+1}&=\frac{\Delta s_k}{36(\Delta s_{k-1}+\Delta s_k)(\Delta s_{k-2}+\Delta s_{k-1}+\Delta s_k)}\\
&\cdot\Bigl[(3\Delta s_k+4\Delta s_{k-1}+2\Delta s_{k-2})^2\\
&+2(\Delta s_{k-1}+2\Delta s_{k-2})(\Delta s_{k-1}-\Delta s_{k-2})\Bigr]\,.
\end{split}
\end{align*}}
\section{Numerical derivatives for the cubic Hermitian method}\label{appendix:A}
Section~\ref{sec:sec4} anticipates that second-order accurate numerical derivatives for $\mathbf K$ and $\boldsymbol{\epsilon}$ are sufficient to maintain fourth-order accuracy with the cubic Hermitian method. Without loss of generality, one assumes a purely absorbing medium, i.e., $\boldsymbol{\epsilon}=0$. In the local truncation error analysis the numerical values of the propagation matrix are considered as exact, namely $\mathbf K(s_k)=\mathbf K_k$ and $\mathbf K(s_{k+1})=\mathbf K_{k+1}$, and one assumes that $\mathbf I(s_k)=\mathbf I_k$. Let the Stokes vector be three times differentiable, allowing its third-order Taylor expansion
{\small
\begin{align*}
\mathbf I(s_{k+1}) = \mathbf I(s_k) + h\mathbf I'(s_k) + h^2\mathbf I''(s_k)/2 + h^3\mathbf I'''(s_k)/6 + O(h^4)\,,
\end{align*}}\noindent
where, for notational simplicity, one denotes $\Delta s=h$. Moreover, let the propagation matrix be twice differentiable, allowing the following Taylor expansions
\begin{align*}
\mathbf K(s_{k+1}) &= \mathbf K(s_k) + h\mathbf K'(s_k) + h^2\mathbf K''(s_k)/2 + O(h^3)\,,\\
\mathbf K'(s_{k+1})&=\mathbf K'(s_k)+h\mathbf K''(s_k)+O(h^2)\,.
\end{align*}
Next, one inserts these Taylor expansions in the analytical homogeneous version of Equation~\eqref{recursive_hermite}, namely
\begin{equation*}
\tilde{\mathbf{\Phi}}_{k+1}\mathbf I(s_{k+1})=\tilde{\mathbf{\Phi}}_k\mathbf I(s_k)\,,
\end{equation*}
with
{\small
\begin{align*}
\tilde{\mathbf{\Phi}}_k&=\mathbf{1}-\frac{h}{2}\mathbf K(s_k)+\frac{h^2}{12}\left[\mathbf K(s_k)\mathbf K(s_k)-\mathbf K'(s_k)\right]\,,\\
\tilde{\mathbf{\Phi}}_{k+1}&=\mathbf{1}+\frac{h}{2}\mathbf K(s_{k+1})+\frac{h^2}{12}\left[\mathbf K(s_{k+1})\mathbf K(s_{k+1})-\mathbf K'(s_{k+1})\right]\,.
\end{align*}}\noindent
Making use of the identities
{\small
\begin{align*}
\mathbf I'(s_k)&= -\mathbf K(s_k)\mathbf I(s_k)\,,\\
\mathbf I''(s_k)&= -\mathbf K'(s_k)\mathbf I(s_k)-\mathbf K(s_k)\mathbf I'(s_k)\,,\\
\mathbf I'''(s_k)&= -\mathbf K''(s_k)\mathbf I(s_k)-2\mathbf K'(s_k)\mathbf I'(s_k)-\mathbf K(s_k)\mathbf I''(s_k)\,,
\end{align*}}\noindent
one performs some algebraic manipulation. The cancellation of all the terms until third-order in $h$ indicates the method as fourth-order accurate. It must be stressed that the first derivative of the propagation matrix is only first-order Taylor-expanded. Second-order accuracy in the numerical derivatives for $\mathbf K$ indicates that
{\small
\begin{align*}
\mathbf K'_k-\mathbf K'(s_k)&=O(h^2)\,\Rightarrow\,\mathbf K'_k=\mathbf K'(s_k)+O(h^2)\\
\mathbf K'_{k+1}\!-\!\mathbf K'(s_{k+1})&\!=\!O(h^2)\Rightarrow\mathbf K'_{k+1}\!=\!\mathbf K'(s_k)\!+\!h\mathbf K''(s_k)\!+\!O(h^2)\,,
\end{align*}}\noindent
introducing only second-order perturbations. Therefore, second-order accurate numerical approximations for $\mathbf K'_k$ and $\mathbf K'_{k+1}$ are found to be sufficient to maintain fourth-order accuracy in the cubic Hermitian method.
%
\bibliographystyle{apj} 
\bibliography{bibfile2}
\end{document}